# Technical Report

27 June 2019

*Engineering study on the use of Head-Mounted display for Brain-Computer Interface*

~

A. Andreev, G. Cattan, M. Congedo

GIPSA-lab, CNRS, University Grenoble-Alpes, Grenoble INP.
Address : GIPSA-lab, 11 rue des Mathématiques, Grenoble Campus BP46, F-38402, France

**Abstract** – In this article we explore the availability of head-mounted display (HMD) devices which can be coupled in a seamless way with P300-based brain-computer interfaces (BCI) using electroencephalography (EEG). The P300 is an event-related potential appearing about 300ms after the onset of a stimulation. The recognition of this potential on the ongoing EEG requires the knowledge of the *exact* onset of the stimuli. In other words, the stimulations presented in the HMD must be perfectly synced with the acquisition of the EEG signal. This is done through a process called *tagging*. The tagging must be performed in a reliable and robust way so as to guarantee the recognition of the P300 and thus the performance of the BCI. An HMD device should also be able to render images fast enough to allow an accurate perception of the stimulations, and equally to not perturb the acquisition of the EEG signal. In addition, an affordable HMD device is needed for both research and entertainment purposes. In this study we selected and tested two HMD configurations.

**Résumé** – Dans cet article, nous recherchons un casque de réalité virtuelle compatible avec l'utilisation des interfaces-cerveau ordinateurs reposant sur l'électroencéphalographie (EEG) et l'exploitation du P300. Le P300 est un potentiel évoqué apparaissant dans l'EEG environ 300ms après le début d'une stimulation. L'exploitation de ce potentiel dans l'EEG nécessite de connaître avec *précision* le début des stimulations. En d'autres termes, le matériel de réalité virtuelle doit être parfaitement synchronisé avec l'acquisition du signal EEG, ce qui est réalisé grâce à un processus appelé *tagging*. Le matériel de réalité virtuelle doit permettre un tagging robuste et précis afin de s'assurer de la bonne détection du P300, laquelle détermine la performance de la BCI. En plus d'avoir un prix accessible, le casque doit également simuler des environnements de qualité, de sorte que les stimuli soient correctement perçus par l'utilisateur, et dans le même temps garantir la qualité du signal EEG. Dans cette étude, nous choisissons et testons deux configurations possibles.

**Introduction**

In recent years virtual reality (VR) has become a popular entertainment method. At the same time, new algorithms for brain-computer interface (BCI) have also been developed allowing for faster bits/min communication (1). The coupling of VR and BCI has been done before, as in (2), but there are numerous engineering challenges that must be addressed since they affect the performance of a VR+BCI interface. The market offers a wide range of devices for virtual reality (see non-exhaustive classification in **Figure 1**). For example, we could use a CAVE (Cave Automatic Virtual Environment) device (3) but these devices are bulky, they take a lot of space and are not suitable for the general public. The first group of devices that can be used in a home environment are of type Linked-to-PC devices (such as Oculus – Facebook, Menlo Park, US - and HTC Vive - HTC, Taoyuan, Taiwan) which are usually connected to a PC through a cable and where all the software and rendering are executed on a powerful computer. The second group are Mobile Head-Mounted Devices (HMD), which are not connected to a desktop computer. These can be separated into Active HMDs, which contain some electronics (such as the SamsungGear – Samsung, Seoul, South Korea – or the Oculus Quest – Facebook, Menlo Park, US) and Passive HMD (such as Google Cardboard – Google, Mountain View, US), which do not contain any electronics. A mobile HMD is compounded by a mask in which we insert a smartphone. Hence, a mobile HMD may be constituted by a large number of masks and smartphones, although they are not always compatible. In this article we will argue against choosing either a Linked-to-PC headset or an active HMD. Then, we will explain why we suggest the use of the VRElegiant (Elegiant, Austin, US) mask. Finally, we will present testing results of using two smartphones, the Huawei mate 7 (Huawei, Shenzhen, China) and the Samsung S6 (Seoul, South Korea), suggesting preference for the former.

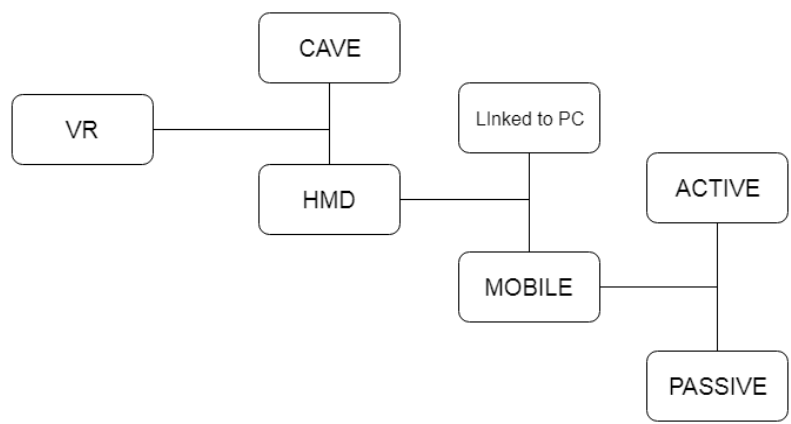

**Figure 1.** Classification of VR devices

**Why Linked-to-PC headset and Mobile Active HMD are not suitable**

Linked-to-PC headsets sometimes include a positional tracking system through external or internal camera (*outside-in* and *inside-out* positional tracking, respectively). This technology is used to precisely locate the user in the real world, as well as to correct the drift due to the inertial measurement unit (IMU). This drift appears because of the hardware imprecision of the IMU, causing the virtual world to slowly move around the user (4–6). The first version of the Oculus (Facebook, Menlo Park, US) and the HTC Vive (HTC, Taoyuan, Taiwan) are the main devices using external captors to track the user's position, however such technology tends to be replaced by inside-out tracking, like the Daydream framework (Google, Mountain View, US), among others. Inside-out tracking is based on *sensor fusion* algorithms (*e.g.*, (7–9)), which basically include the camera input to correct the IMU drift. In some ways, this technology is immature and faces several problems such as: the restriction of the field of view (*e.g.*, HoloLens - Microsoft, Redmond, US), the loss of virtual object or user position and the restriction of the game area (*e.g.*, Lenovo Mirage – Lenovo, Beijing, China) or the non-continuous updating of this area, that is, the incapacity of detecting events in the real world (*e.g.*, Occipital Bridge - Occipital, San Francisco, US) – an important aspect of augmented reality applications, for example. At this point, the Oculus Quest (Facebook, Menlo Park, US) is probably the only *effective* Mobile Active HMD for VR, based on this technology.

Although positional tracking can add to the user's experience, Linked-to-PC VR devices also have several disadvantages such as the use of: a cable, which restricts the movement and a powerful computer which is cumbersome and expensive. On the other hand, even though Mobile Active HMD have the advantage of integrated IMU, they are often proprietary and more expensive as compared to Mobile Passive HMDs. In addition the effect of the electronics in an Active HMD on the EEG signal has not been studied at the time when this study has been performed.

**Why the VRElegiant**

We suggest the use of a Mobile Passive HMD, consisting of a mask with no electronics and a regular smartphone, that is, a potential *ubiquitous* VR technology, enabling the widespread diffusion of VR+BCI applications. Among these masks, the VRElegiant (**Figure 2**) headset (Elegiant, Austin, US) is affordable, comfortable and adapts to a wide range of smartphones.

It is also easy to add an extra USB cable for our evaluation. VRElegiant has a field of view about 96 degrees.

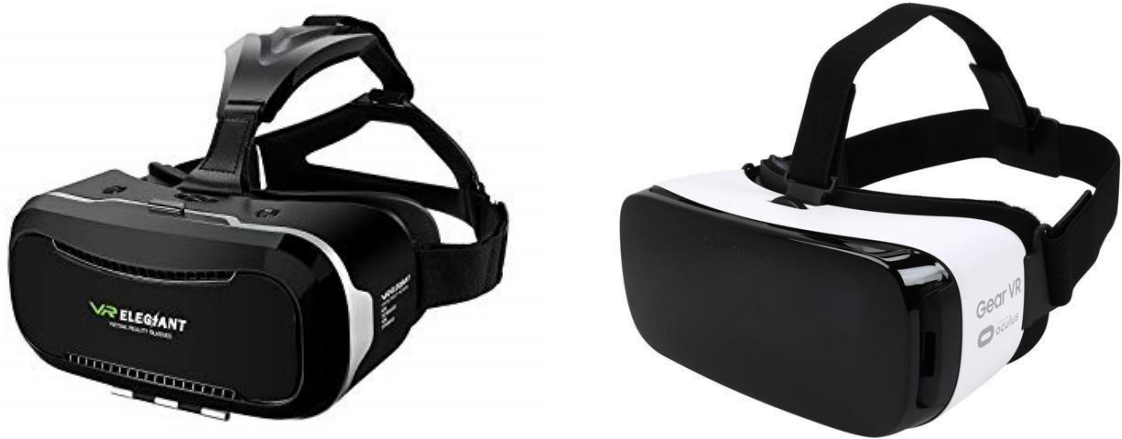

**Figure 2.** VRElegiant Passive HMD on the left and Samsung Gear Active HMD on the right

**Evaluation**

Visually P300-based BCIs require to tag the EEG with the exact onset of the visual stimuli appearing on the screen. Incorrectly marking the onset of the stimulations on the EEG time line induces *latency* and *jitter*. The latency is the average delay between the tagging of these onsets on the EEG and the actual visualisation of these stimuli by the user. The jitter is the standard deviation of this latency, meaning the inter-trial variability of these delays. Therefore, a low jitter communication between the application running on the phone inside the HMD and the EEG acquisition system is necessary. Thus we suggest using hardware communication through USB between the two as in (10). This method is known as USB tagging and we have successfully implemented it within the Huawei Mate 7 (Huawei, Shenzhen, China) coupled with VRElegiant and the Samsung S6 (**Figure 3**) smartphone coupled with SamsungGear headset (Samsung, Seoul, South Korea), which can also be used as a Mobile Passive HMD. The Huawei Mate 7 is a middle-range smartphone, affordable for the general public. It also has a large screen 1920 x 1080 (386 ppi[1]), which makes it very interesting to improve the immersion feeling in VR. The Samsung S6 is a high-end device with a better display and resolution of 2560 x 1440 (577 ppi).

---

[1] Pixel per inch

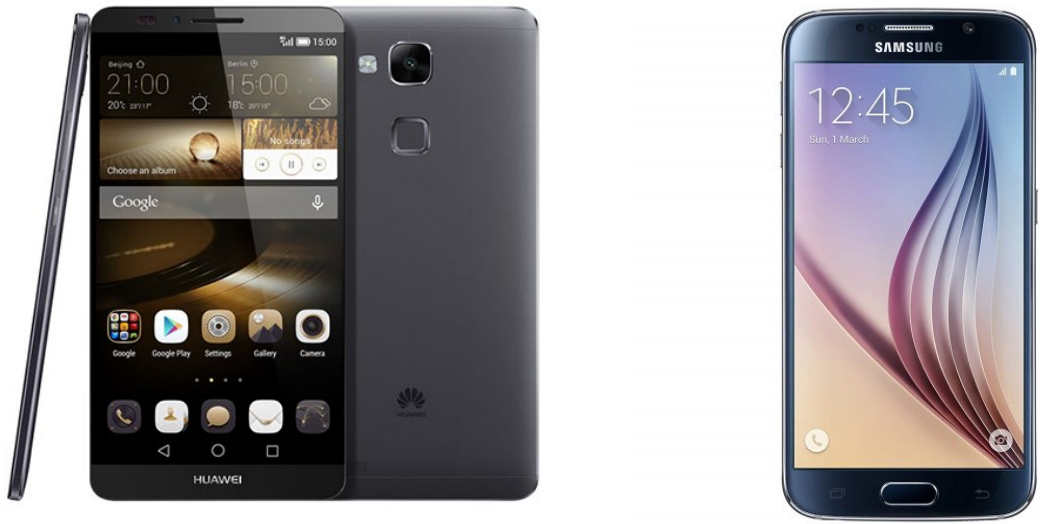

**Figure 3.** Left Huawei Mate 7, right Samsung S6

However, since the USB port was not available on the SamsungGear, we explored another method of communication which consists of flashing the smartphone torchlight at each stimulus. The resulting physical flash is then converted to a numeric tag thanks to a photodiode connected to the EEG acquisition system. In practice, this method is less robust than the USB tagging since there is only two possible values (flash or no-flash) and it is also prone to losing some of the flashes. It also introduces a higher jitter in comparison to the USB tagging under PC or VR (**Figure 4**).

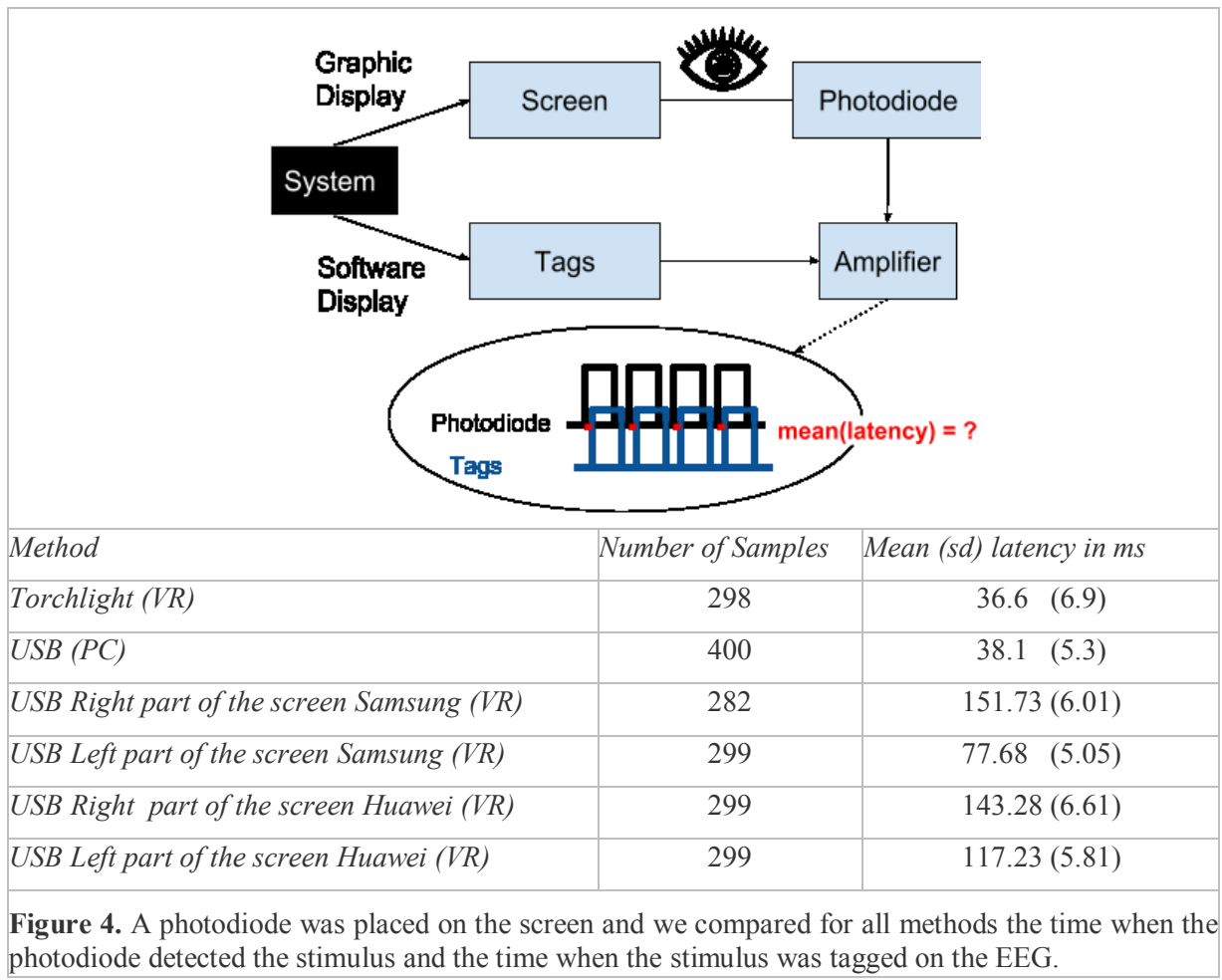

| Method | Number of Samples | Mean (sd) latency in ms |
|---|---|---|
| *Torchlight (VR)* | 298 | 36.6 (6.9) |
| *USB (PC)* | 400 | 38.1 (5.3) |
| *USB Right part of the screen Samsung (VR)* | 282 | 151.73 (6.01) |
| *USB Left part of the screen Samsung (VR)* | 299 | 77.68 (5.05) |
| *USB Right part of the screen Huawei (VR)* | 299 | 143.28 (6.61) |
| *USB Left part of the screen Huawei (VR)* | 299 | 117.23 (5.81) |

**Figure 4.** A photodiode was placed on the screen and we compared for all methods the time when the photodiode detected the stimulus and the time when the stimulus was tagged on the EEG.

The photodiode represents our baseline as it plays the role of a human subject who perceives the visual stimulation used in a visual P300-based BCI. Both the photodiode and USB (or torchlight) tagging have been recorded and synced by an EEG amplifier as USB to parallel port tags and channel 1 of the amplifier, respectively. The amplifier used is a 16-channel g.USBamp (GTEC, New York, US) using a frequency of 512 Hz and the recoding software is OpenVibe (11,12). In VR the screen is split in two parts, each part rendering a texture for a different eye. This might result in the perception of two stimuli if the inter-oculus latency is high. Therefore, we requested the smallest possible difference between the two screens. In general, a high mean latency is acceptable if the inter-oculus latency is small, as it is possible to shift the EEG signal by a fixed time interval computed on the basis of this mean latency (13). However, this would not solve the problem in the case of a high jitter. A number of technical improvements have been applied in order to decrease the jitter. The smartphones were switched in air-plane mode, the asynchronous messages were disabled in the code implementation, the texture resolution was divided by 8, the antialiasing was suppressed, the texture distortion shader was removed and the pre- and post-rendering operations were factorized when they were duplicated between right and left parts of the screen. We observed that these actions divided the jitter from 11.88 (average over 299 flashes) to 6.01 (average over 282 flashes) for the right part of the screen of the Samsung S6. Surprisingly, the comparison of the two smartphones showed that the Huawei mate 7 has the smallest difference between screens (**Figure 4**), from which our preference for this device.

**Discussion and Conclusion**

We have studied the usability of VR devices for P300-based BCI and in particular jitter communication. For our tests we have selected affordable devices with preference for Mobile Passive HMD. Our tests show that in VR mode even after a number of technical optimizations we have a considerable delay as compared to PC mode. Also, because of the nature of rendering in VR mode (the splitting of the screen) all further tests need to consider testing on both sides (each one dedicated to different eye) in order to have a realistic evaluation of the jitter. The jitter varies among different devices thus these aspects have to be studied as well for Active and Linked-To-PC HMD.

We would also like to note that even though hardware USB tagging in Mobile HMD gives better results than, for example, Wi-Fi tagging and it is more robust than torchlight, it has the disadvantage of using an extra USB cable. Wi-Fi tagging needs to be studied further as a method which puts timestamp in the tags and synchronises the clocks of the amplifier and the smartphone at regular intervals has good potential**.** We were unable to test this as the g.USBamp does not provide timestamps for the samples of recorded signal. However, this functionality may be adapted from existing implementations such as (14,15). These considerations also stand for Active Mobile HMD devices.

A recent study (16) shows that the EEG signal is not affected when using a Mobile Passive HMD which is very encouraging. The future of the mixed technology VR+BCI looks bright and recommendations on the use of such technology in gaming can be found in (17). Also a number of low cost with improved quality EEG mobile devices such as OpenBCI (New York, US) have been developed. In this vein, research is ongoing to provide research grade, affordable and open EEG (http://eeg.io). These headsets, combined with low cost HMD devices can pave the way for wider spread of VR+BCI applications.

# References

1. Congedo M, Barachant A, Bhatia R. Riemannian geometry for EEG-based brain-computer interfaces; a primer and a review. Brain-Comput Interfaces. 2017;4(3):155–74.

2. Käthner I, Kübler A, Halder S. Rapid P300 brain-computer interface communication with a head-mounted display. Front Neurosci. 2015;9:207.

3. Cruz-Neira C, Sandin DJ, DeFanti TA, Kenyon RV, Hart JC. The CAVE: Audio Visual Experience Automatic Virtual Environment. Commun ACM. 1992 Jun;35(6):64–72.

4. Safaeifar A, Nahvi A. Drift cancellation of an orientation tracker for a virtual reality head-mounted display. In: 2015 3rd RSI International Conference on Robotics and Mechatronics (ICROM). 2015. p. 296–301.

5. Latt WT, Veluvolu KC, Ang WT. Drift-Free Position Estimation of Periodic or Quasi-Periodic Motion Using Inertial Sensors. Sensors. 2011 May 31;11(6):5931–51.

6. Burdea GC, Coiffet P. Virtual Reality Technology. John Wiley & Sons; 2003. 472 p.

7. Hol JD, Schon TB, Gustafsson F, Slycke PJ. Sensor Fusion for Augmented Reality. In: 2006 9th International Conference on Information Fusion. 2006. p. 1–6.

8. Sensor fusion algorithm [Internet]. Available from: http://www.google.com/patents/US8548608

9. Perek DR, Schwager MA, Drasnin S, Seilstad MJ. Sensor fusion algorithm [Internet]. US8548608B2, 2013. Available from: https://patents.google.com/patent/US8548608B2/en

10. Andreev A, Barachant A, Lotte F, Congedo M. Recreational Applications of OpenViBE: Brain Invaders and Use-the-Force [Internet]. Vol. chap. 14. John Wiley ; Sons; 2016. Available from: https://hal.archives-ouvertes.fr/hal-01366873/document

11. Renard Y, Lotte F, Gibert G, Congedo M, Maby E, Delannoy V, et al. OpenViBE: An Open-Source Software Platform to Design, Test, and Use Brain–Computer Interfaces in Real and Virtual Environments. Presence Teleoperators Virtual Environ. 2010 Feb 1;19(1):35–53.

12. Arrouët C, Congedo M, Marvie J-E, Lamarche F, Lécuyer A, Arnaldi B. Open-ViBE: A Three Dimensional Platform for Real-Time Neuroscience. J Neurother. 2005 Jul 8;9(1):3–25.

13. Cattan G, Andreev A, Maureille B, Congedo M. Analysis of tagging latency when comparing event-related potentials [Internet]. Grenoble: Gipsa-Lab ; IHMTEK; 2018 Dec. (GIPSA-VIBS). Available from: https://hal.archives-ouvertes.fr/hal-01947551

14. Stenner T, Boulay C, Medine D. LabStreamingLayer [Internet]. Swartz Center for Computational Neuroscience; 2015. Available from: https://github.com/sccn/labstreaminglayer

15. Foy N. TCP Tagging (Software Tagging) [Internet]. OpenViBE. 2016. Available from: http://openvibe.inria.fr/tcp-tagging/

16. Cattan G, Andreev A, Mendoza C, Congedo M. The Impact of Passive Head-Mounted Virtual Reality Devices on the Quality of EEG Signals. In Delft: The Eurographics Association; 2018. Available from: http://dx.doi.org/10.2312/vriphys.20181064

17. Cattan G, Mendoza C, Andreev A, Congedo M. Recommendations for Integrating a P300-Based Brain Computer Interface in Virtual Reality Environments for Gaming. Computers. 2018 May 28;7(2):34.